\documentstyle{article}

\newcommand{\ba}{\begin{eqnarray}}
\newcommand{\ea}{\end{eqnarray}}
\input{psfig}
\begin{document}
\pagestyle{plain}
\def\ii{\'\i}

\title{Strong decays of nucleon and delta resonances}
\author{R.~Bijker\\
Instituto de Ciencias Nucleares, U.N.A.M.,\\
Apartado Postal 70-543, 04510 M\'exico D.F., M\'exico
\and
A.~Leviatan\\
Racah Institute of Physics, The Hebrew University,\\
Jerusalem 91904, Israel}
\maketitle

\begin{center}
\end{center}

\begin{abstract}
We study the strong couplings of the nucleon and delta resonances 
in a collective model. In the ensuing algebraic treatment we derive 
closed expressions for decay widths which are used to analyze the 
experimental data for strong decays into the pion and eta channels.
\end{abstract}

\clearpage

\section{Introduction}

A large amount of experimental data was accumulated in the 1960's 
and 1970's on the spectroscopy of light hadrons. These data led first 
to the introduction of $SU_f(3)$ by Gell'Mann \cite{MGM} and Ne'eman 
\cite{YN} (later enlarged to $SU_{sf}(6)$ by G\"ursey and Radicati 
\cite{GR}), and subsequently to $SU_c(3)$ color symmetry as a gauge 
symmetry of strong interactions. The construction of dedicated 
facilities ({\it e.g.} CEBAF, MAMI, ELSA) that promise to produce new  
and more accurate data, have stimulated us to reexamine baryon 
spectroscopy. 

In our reanalysis we have introduced, in addition to the basic 
spin-flavor-color symmetry, $SU_{sf}(6) \otimes SU_c(3)$, a new 
ingredient, namely a space symmetry, ${\cal G}$, which was taken to be 
${\cal G}=U(7)$ for baryons \cite{BIL}. The space symmetry makes it
possible to study, in a straightforward way, several limiting situations 
({\it e.g.} harmonic oscillator and collective dynamics), and to derive 
a set of transparent results that can be used to analyze the experimental 
data. This algebraic approach has been used \cite{BIL,emff}
to analyze the mass spectrum and electromagnetic couplings of 
nonstrange baryon resonances. It presents an alternative for 
the use of nonrelativistic or relativized Schr\"odinger equations 
\cite{IK,CI}. In addition to electromagnetic couplings, strong decays 
of baryons provide an important, complementary, tool to study the 
structure of baryons. 

The strong decays have been analyzed previously in the 
nonrelativistic \cite{KI} and relativized quark models \cite{CR}. 
These models emphasize single-particle aspects of quark dynamics 
in which only a few low-lying configurations in the confining potential 
contribute significantly to the baryon wave function. In the framework 
of the earlier mentioned algebraic approach it is possible to study 
also other, more collective, types of dynamics. In this contribution, 
we investigate in detail the strong decays in a collective model of 
baryon structure. 
 
\section{Algebraic model of the nucleon}

\begin{figure}
\centerline{\hbox{
\psfig{figure=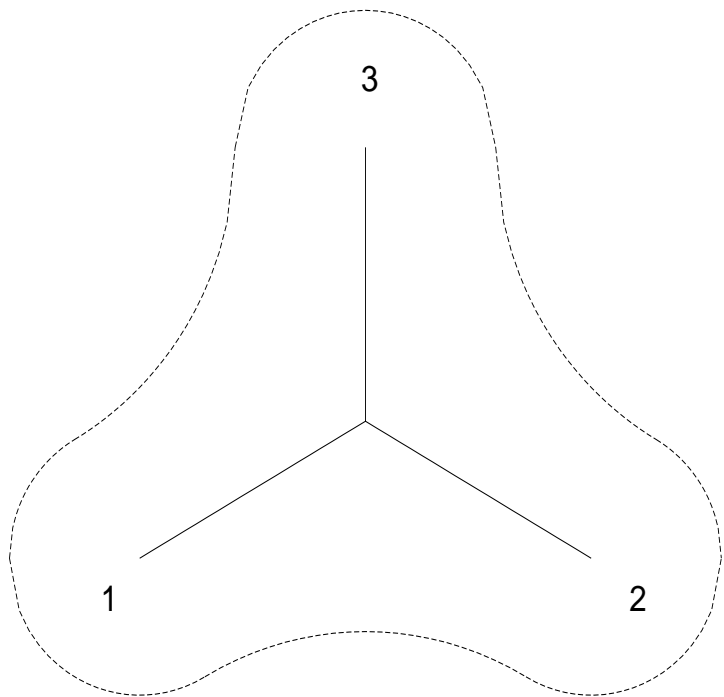,height=0.65\textwidth,width=0.8\textwidth} }}
\caption{
Collective model of baryons.
\label{geometry}}
\end{figure}

In \cite{BIL} we introduced an algebraic model, in which the nucleon 
has the string configuration of Figure~\ref{geometry}. Its three 
constituent parts are characterized by the internal degrees of freedom 
of spin, flavor and color and by the two relative Jacobi coordinates, 
$\vec{\rho}$ and $\vec{\lambda}$, and their conjugate momenta. For these 
six spatial degrees of freedom we suggested to use a $U(7)$ spectrum 
generating algebra whose building blocks are six dipole
bosons, $b^{\dagger}_{\rho,i}$ and $b^{\dagger}_{\lambda,i}$ 
($i=x,y,z$), and an auxiliary scalar boson, $s^{\dagger}$. 
For a system of interacting bosons the model space is spanned by the 
symmetric irreps $[N]$ of $U(7)$, which contains the oscillator shells 
with $n=0,1,2,\ldots,N$. Here $N$ is the conserved total number of bosons. 
We note that the scalar boson does not introduce a new degree of freedom, 
since for a given total boson number $N$ it can always be eliminated by 
$s \rightarrow \sqrt{N-\hat n_{\rho}-n_{\lambda}}$ (Holstein-Primakoff 
realization of $U(7)$). 
The full algebraic structure is obtained by combining the geometric part, 
$U(7)$, with the internal spin-flavor-color part, 
$SU_{sf}(6) \otimes SU_c(3)$. 

For the nucleon (isospin $I=1/2$) and delta ($I=3/2$) families of 
resonances the three strings of Figure~\ref{geometry} have equal
length and equal relative angles. Hence this configuration is an 
oblate top and has $D_{3h}$ point group symmetry.
The classification under $D_{3h}$ is equivalent to
the classification under permutations and parity.
States are characterized by 
$(v_1,v_2);K,L_t^{P}$, where $(v_1,v_2)$ denote the vibrations
(stretching, bending); $K$ denotes the projection of the rotational
angular momentum $L$ on the body-fixed symmetry axis, $P$ the parity and
$t$ the symmetry type of the state under $D_3$ (a subgroup of $D_{3h}$ 
isomorphic to the $S_3$ permutation group).
The symmetry type of the geometric part must be the same as that 
of the spin-flavor part (the color part is antisymmetric). Therefore,
one can use the representations of either $D_3$ or $SU_{sf}(6)$ 
to label the states: $A_1 \leftrightarrow 56$,
$A_2 \leftrightarrow 20$, $E \leftrightarrow 70$.
We use the latter notation and denote the total baryon wave function as
\ba
\left| \, ^{2S+1}\mbox{dim}\{SU_f(3)\}_J \, 
[\mbox{dim}\{SU_{sf}(6)\},L^P]_{(v_1,v_2);K} \, \right> ~,
\ea
where $S$ and $J$ are the spin and total angular momentum 
$\vec{J}=\vec{L}+\vec{S}$.
In this collective model of the nucleon baryon resonances are interpreted 
as vibrations and rotations of an oblate symmetric top.
The corresponding wave functions, when expressed in a harmonic oscillator
basis, are spread over many shells and hence are truly collective. 

\section{Strong couplings}

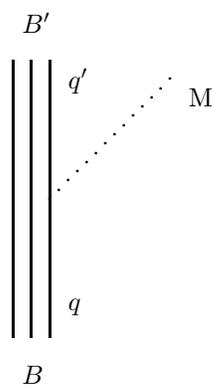
\begin{figure}
\centering
\setlength{\unitlength}{0.7pt}
\begin{picture}(200,250)(0,-25)
\thicklines
\put ( 25, 25) {\line (0,1){150}}
\put ( 35, 25) {\line (0,1){150}}
\put ( 45, 25) {\line (0,1){150}}
\put ( 30,  0) {$B$}
\put ( 30,190) {$B^{\prime}$}
\put ( 55, 40) {$q$}
\put ( 55,160) {$q^{\prime}$}
\multiput ( 45,100)( 5, 5){14}{\circle*{0.1}}
\put (120,150) {M}
\end{picture}
\caption[]{\small 
Elementary meson emission.
\normalsize} 
\label{qqM}
\end{figure}

In this contribution we study strong decays of baryons 
by the emission of a pseudoscalar meson 
\ba 
B \rightarrow B^{\prime} + M ~.
\ea
In order to calculate the decay widths of this process we have to specify 
the form of the operator inducing the transition. Several forms have been 
discussed in \cite{LeYaouanc}. Here we use the transition operator that 
has also been used in \cite{KI}
\ba
{\cal H} &=& \frac{1}{(2\pi)^{3/2} (2k_0)^{1/2}} 
\sum_{j=1}^{3} X^M_{j} \left[ 
2g \, (\vec{s}_j \cdot \vec{k}) \mbox{e}^{-i \vec{k} \cdot \vec{r}_j}
+ h \, \vec{s}_j \cdot 
(\vec{p}_j \, \mbox{e}^{-i \vec{k} \cdot \vec{r}_j} +
\mbox{e}^{-i \vec{k} \cdot \vec{r}_j} \, \vec{p}_j) \right] ~, \label{hs}
\ea
where $\vec{r}_j$, $\vec{p}_j$ and $\vec{s}_j$ are the coordinate, 
momentum and spin of the $j$-th constituent, respectively; 
$k_0=E_M=E_B-E_{B^{\prime}}$ is the meson energy, and 
$\vec{k}=\vec{P}_M=\vec{P}-\vec{P}^{\prime}=k \hat z$ denotes the 
momentum carried by the outgoing meson. Here $\vec{P}=P_z \hat z$
and $\vec{P}^{\prime}$ ($=P^{\prime}_z \hat z$) are the momenta of the 
initial and final baryon. The flavor operator $X^M_j$ (expressed in 
terms of Gell-Mann matrices) corresponds to the 
emission of an elementary meson by the $j$-th constituent: 
$q_j \rightarrow q_j^{\prime} + M$ (see Figure~\ref{qqM}).
The coefficients $g$ and $h$ are parameters.

Using the symmetry of the wave functions for nonstrange baryons, 
transforming to Jacobi coordinates and integrating over the baryon 
center of mass coordinate, we find
\ba
{\cal H} &=& \frac{1}{(2\pi)^{3/2} (2k_0)^{1/2}} \, 
6 X^M_{3} \left[ g \, k s_{3,z} \, \hat U 
- h \, s_{3,z} (\hat T_z - \frac{1}{6}(P_z+P^{\prime}_z) \hat U)
- \frac{1}{2} h \, (s_{3,+} \hat T_- + s_{3,-} \hat T_+) \right] ~.
\nonumber\\
\label{hstrong}
\ea
The operators $\hat U$ and $\hat T_+$ only act on the spatial part 
of the baryon wave function. In an algebraic treatment they are 
given by 
\ba
\hat U &=& \mbox{e}^{ i k \beta \hat D_{\lambda,z}/X_D } ~,
\nonumber\\
\hat T_m &=& -\frac{i m_{3} k_0 \beta}{2 X_D} \left( \hat D_{\lambda,m} \,
\hat U + \, \hat U \, \hat D_{\lambda,m} \right) ~, \label{emop}
\ea
where $\hat D_{\lambda,m} = (b^{\dagger}_{\lambda} \times s -
s^{\dagger} \times \tilde b_{\lambda})^{(1)}_m$ is a dipole operator 
in $U(7)$ and has the same transformation properties as the Jacobi 
coordinate $\lambda_m$. The coefficient $X_D$ is a normalization 
factor and $\beta$ represents the scale of the coordinate \cite{BIL}. 

Since $\hat D_{\lambda}$ is a generator of the algebra of $U(7)$, 
the matrix elements of $\hat U$ are representation 
matrix elements of $U(7)$, {\it i.e.} generalized Wigner 
${\cal D}$-matrices. By making an appropriate 
basis transformation they can be obtained exactly. In addition, in 
the limit of $N \rightarrow \infty$ (large model space)
they can also be derived in closed form. This derivation consists of 
several steps. The rotational states $|K,L,M \rangle$ belonging to the 
the vibrational ground state band of the oblate symmetric top 
with $(v_1,v_2)=(0,0)$, can be obtained by projection from an intrinsic 
(or coherent) state 
\ba
|K,L,M\rangle &=& \sqrt{\frac{2L+1}{8\pi^2}} \int d\Omega \,
{\cal D}_{MK}^{(L) \ast}(\Omega) \, {\cal R}(\Omega) \, |N,R\rangle ~,
\nonumber\\
|N,R\rangle &=& \frac{1}{\sqrt{N!(1+R^2)^N}} 
\left[ s^{\dagger} + \frac{R}{\sqrt{2}} (b^{\dagger}_{\lambda,x}
+b^{\dagger}_{\rho,y})\right]^N \, |0\rangle ~. \label{klm}
\ea
The coefficient $R$ appears in the mass operator and is associated 
with the size of the string \cite{BIL}. 
Next we construct states with good $D_3$ symmetry by taking 
the linear combinations 
\ba
|\psi_1\rangle &=& \frac{1}{\sqrt{2(1+\delta_{K,0})}} \,
\left[ (-)^L |K,L,M\rangle + |-K,L,M\rangle \right] ~,
\nonumber\\
|\psi_2\rangle &=& \frac{i}{\sqrt{2(1+\delta_{K,0})}} \,
\left[ |K,L,M\rangle -(-)^L |-K,L,M\rangle \right] ~. \label{wf}
\ea
For $K(\mbox{mod }3)=0$ the wave function $|\psi_1\rangle$ is symmetric 
($A_1 \leftrightarrow 56$) and $|\psi_2\rangle$ antisymmetric 
($A_2 \leftrightarrow 20$), whereas for $K(\mbox{mod }3) \neq 0$ 
the wave functions $|\psi_1\rangle$ and $|\psi_2\rangle$ are the 
$E_{\lambda}$ and $E_{\rho}$ components, respectively, 
of the mixed-symmetry doublet ($E \leftrightarrow 70$).
Eq.~(\ref{wf}) is consistent with the choice of geometry in $|N,R\rangle$.
The matrix element of $\hat U$ relevant for a decay in which the final 
state is either the nucleon or the delta with 
$| \psi_0 \rangle = |K=0,L=0,M=0 \rangle$ is given by
\ba
\langle \psi_0 | \hat U | \psi_1 \rangle &=& \delta_{M,0} \, F(k\beta) ~,
\ea
with
\ba
F(k\beta) &=& i^{K} \, \sqrt{\frac{2L+1}{2(1+\delta_{K,0})}} 
\frac{\sqrt{(L-K)!(L+K)!}}{(L-K)!!(L+K)!!} \,
\left[ 1+(-1)^{L-K} \right] \, j_{L}(k\beta) ~.
\ea
In the derivation 
we have used that in the large $N$ limit the intrinsic matrix 
element becomes diagonal in the orientation $\Omega$ of the condensate.
For the the matrix elements of $\hat T$ we find similar expressions 
in terms of spherical Bessel functions 
\ba
\langle \psi_0 | \hat T_z | \psi_1 \rangle 
&=& - \delta_{M,0} \, m_3 k_0 \beta \, \frac{d F(k\beta)}{d (k\beta)} ~,
\nonumber\\
\langle \psi_0 | \hat T_{\pm} | \psi_1 \rangle
&=& \mp \delta_{M,\mp 1} \, m_3 k_0 \beta \, \sqrt{L(L+1)} \, 
\frac{F(k\beta)}{k\beta} ~.
\ea
The operators $\hat U$ and $\hat T_m$ do not connect the nucleon 
and delta wave function $\psi_0$ with $\psi_2$ of Eq.~(\ref{wf}).
In the collective model discussed here, these spatial matrix elements 
are folded with a distribution function 
\ba
g(\beta) &=& \beta^2 \mbox{e}^{-\beta/a}/2a^3 ~. \label{gbeta}
\ea 
With this distribution function we reproduce the observed dipole form 
for the electric form factor of the proton $G^p_E=1/(1+k^2a^2)^2$~. 
In Table~\ref{collff} we present some collective 
matrix elements ${\cal F}(k)=\int d\beta \, g(\beta) \, F(k\beta)$ 
for the decay of resonances which are interpreted in the collective 
model as rotational excitations and have $(v_1,v_2)=(0,0)$. 
We note that the matrix elements of $\hat U$ and $\hat T$ between 
states belonging to the $(v_1,v_2)=(0,0)$ ground state band do not 
depend on the coefficient $R$ of Eq.~(\ref{klm}). The decays involving 
resonances associated with 
vibrationally excited states, {\it e.g.} with $(v_1,v_2)=(1,0)$ 
or $(0,1)$, show an explicit dependence on $R$ \cite{BIL,emff}.
In this contribution we only consider resonances that are interpreted 
as rotational excitations and have $(v_1,v_2)=(0,0)$.

The helicity amplitudes for strong decays of baryon resonances 
$B \rightarrow B^{\prime} + M$ can be obtained by combining the spatial 
contribution with the appropriate spin-flavor matrix elements of 
$X^M_3 \, s_{3,m}$. The calculation of the contribution of the 
spin-flavor part is straightforward. 
For the pseudoscalar $\eta$ mesons we introduce, as usual, a  
mixing angle $\theta_P$ between the octet and singlet mesons \cite{PDG} 
\ba
X^{\eta} &=& X^{\eta_8} \, \cos \theta_P - X^{\eta_1} \, \sin \theta_P ~,
\nonumber\\
X^{\eta^{\prime}} &=& X^{\eta_8} \, \sin \theta_P 
+ X^{\eta_1} \, \cos \theta_P ~.
\ea
For decays in which the initial baryon has angular momentum 
$\vec{J}=\vec{L}+\vec{S}$ and in which the final baryon is either the 
nucleon or the delta with $L^{\prime}=0$ and thus
$J^{\prime}=S^{\prime}$, the helicity amplitudes are 
\ba
A_{\nu}(k) &=& \int d \beta \, g(\beta) \, 
\langle \alpha^{\prime},L^{\prime}=0,S^{\prime},J^{\prime}=S^{\prime},\nu \, 
| \, {\cal H} \, | \, \alpha,L,S,J,\nu \rangle ~, \label{anu}
\ea
where $\alpha$ denotes the quantum numbers that, in addition to $L$, $S$, 
$J$ and $\nu$, are needed to specify the baryon wave function.
With the definition of the transition operator in Eq.~(\ref{hs}) and the 
helicity amplitudes in Eq.~(\ref{anu}), the decay widths for a 
specific isospin channel are 
\ba
\Gamma(B \rightarrow B^{\prime} + M) &=& 2 \pi \rho_f \, 
\frac{2}{2J+1} \sum_{\nu>0} | A_{\nu}(k) |^2 ~, \label{dw}
\ea
where $\rho_f$ is a phase space factor.
We adopt the procedure of \cite{LeYaouanc} to calculate the decay widths: 
(i) all calculations are performed in the rest frame of the decaying 
resonance ($P_z=0$ and hence $P_z^{\prime}=-k$), and (ii) 
for the momentum $k$ of the emitted meson and the phase space factor 
$\rho_f$ we use the relativistic expressions 
\ba
k^2 &=& -m_M^2 + \frac{(m_B^2-m_{B^{\prime}}^2+m_M^2)^2}{4m_B^2} ~, 
\nonumber\\
\rho_f &=& 4 \pi \frac{E_{B^{\prime}} E_M k}{m_B} ~.
\ea
Here $E_{B^{\prime}}=\sqrt{m_{B^{\prime}}^2+k^2}$ and 
$E_{M}=\sqrt{m_{M}^2+k^2}$.  

\section{Results}

For all resonances with the same value of $(v_1,v_2),L^P$ 
the expression for the decay widths of Eq.~(\ref{dw}) can be rewritten 
in a more transparent form in terms of only two elementary partial 
wave amplitudes $W_l(k)$ with $l=L \pm 1$, 
\ba
\Gamma(B \rightarrow B^{\prime} + M) 
&=& 2 \pi \rho_f \, \frac{1}{(2\pi)^3 2k_0} \, 
\sum_{l=L \pm 1} c_l \left| W_l(k) \right|^2 ~. \label{width}
\ea
For this set of resonances, the amplitudes $W_l(k)$ contain the $k$ 
dependence, while the coefficients $c_l$ depend on the specific 
baryon resonance involved in the decay. 

In Table~\ref{negwidth} we give the coefficients $c_l$ for 
the negative parity resonances with $(v_1,v_2),L^P=(0,0),1^-$.
In the collective model with distribution given by Eq.~(\ref{gbeta}) 
the corresponding $S$ and $D$ elementary partial wave amplitudes are 
\ba
W_0(k) &=& i \, \left[ 
[gk+\frac{1}{6}h(P_z+P_z^{\prime})] \frac{ka}{(1+k^2a^2)^2}
+h \, m_3k_0a \frac{3-k^2a^2}{(1+k^2a^2)^3} \right] ~,
\nonumber\\
W_2(k) &=& i \, \left[ 
[gk+\frac{1}{6}h(P_z+P_z^{\prime})] \frac{ka}{(1+k^2a^2)^2}
-h \, m_3k_0a \frac{4k^2a^2}{(1+k^2a^2)^3} \right] ~. \label{pw02}
\ea
Partial widths for other models of the nucleon and its resonances 
can be obtained by introducing the corresponding expressions for the 
elementary amplitudes $W_l(k)$. For example, the relevant 
expressions in the harmonic oscillator quark model are 
\ba
W_0(k) &=& \frac{i}{3} \left[ [gk+\frac{1}{6}h(P_z+P_z^{\prime})] k \beta 
+ h m_3 k_0 \beta (3-\frac{k^2 \beta^2}{3}) \right]
\mbox{e}^{-k^2 \beta^2/6} ~,
\nonumber\\
W_2(k) &=& \frac{i}{3} \left[ [gk+\frac{1}{6}h(P_z+P_z^{\prime})] k \beta 
- \frac{1}{3} h m_3 k_0 \beta k^2 \beta^2 \right]
\mbox{e}^{-k^2 \beta^2/6} ~.
\ea

\section{Analysis of experimental data}

We consider strong decays of nucleon and delta resonances into the 
$\pi$ and $\eta$ channels. In Table~\ref{npi} we show a 
comparison between the experimental widths, which are extracted from the 
most recent compilation by the Particle Data Group \cite{PDG}, and 
the results of our calculation. The calculated values depend on the
two parameters $g$ and $h$ in the transition operator of Eq.~(\ref{hs}), 
and on the strong interaction radius $a$ (see Eq.~(\ref{gbeta}). 
The values of $g$, $h$ and $a$ 
are determined in a least square fit to the $N \pi$ partial widths 
(which are relatively well known). The $S_{11}$ resonances were excluded 
from the fit, since for these resonances the situation is not clear due 
to possible mixing of $N(1535)S_{11}$ and $N(1650)S_{11}$ and 
the possible existence of a third $S_{11}$ resonance \cite{ZPLi}. 
As a result we find that the value of the strong interaction radius $a$ 
is the same as the electromagnetic value $a=0.232$ fm \cite{emff}. 
Furthermore $g=1.164$ GeV$^{-1}$ and $h=-0.094$ GeV$^{-1}$. The 
relative sign is consistent with a previous analysis of the strong 
decay of mesons \cite{GIK} and with a derivation from the axial-vector 
coupling (see {\it e.g.} \cite{LeYaouanc}). We note that $g$ and $h$ 
have the {\em same} value for {\em all} resonances and {\em all} decay 
channels. In comparing with previous calculations, it should be noted 
that in the calculation in the nonrelativistic quark model of \cite{KI}  
the decay widths are parametrized by four reduced partial wave ampitudes
instead of the two elementary amplitudes $g$ and $h$. Furthermore 
the momentum dependence of these reduced amplitudes are represented 
by constants. The calculation in the relativized quark model of \cite{CR} 
was done using a pair-creation model for the decay and 
involved a different assumption on the phase space factor. 
Both the nonrelativistic and relativized quark model calculations 
include the effects of mixing induced by the hyperfine interaction,
which in the present calculation are not taken into account.

\begin{figure}
\centering
\setlength{\unitlength}{1pt}
\begin{picture}(300,250)(0,0)
\thicklines
\put (140,  0) {\line(1,0){100}}
\put (  0, 73) {\line(1,0){125}}
\put ( 95,145) {\line(1,0){145}}
\put ( 95,149) {\line(1,0){145}}
\put ( 30,178) {\line(1,0){210}}
\put ( 30,184) {\line(1,0){210}}
\put ( 30,190) {\line(1,0){210}}

\put (150,135) {\line(0,1){55}}
\put (140,122.5) {10(5)}
\put (150,115) {\vector(0,-1){115}}

\put (170, 85) {\line(0,1){99}}
\put (160, 72.5) {72(31)}
\put (170, 65) {\vector(0,-1){65}}

\put (190, 35) {\line(0,1){143}}
\put (175, 22.5) {117(35)}
\put (190, 15) {\vector(0,-1){15}}

\put (210,110) {\line(0,1){39}}
\put (200, 97.5) {79(85)}
\put (210, 90) {\vector(0,-1){90}}

\put (230, 60) {\line(0,1){85}}
\put (215, 47.5) {67(115)}
\put (230, 40) {\vector(0,-1){40}}

\put (130.5,145) {\line(-1,-4){10.5}}
\put (105  , 90.5) {27(12)}
\put (115  , 83) {\vector(-1,-4){ 2.5}}

\put (106.5,149) {\line(-1,-4){ 3.5}}
\put ( 88  ,122.5) {0(23)}
\put ( 98  ,115) {\vector(-1,-4){10.5}}

\put ( 88.75,178) {\line(-1,-4){18.5}}
\put ( 55.25, 91.5) {9(24)}
\put ( 65.25, 84) {\vector(-1,-4){2.75}}

\put ( 65.25,184) {\line(-1,-4){12.5}}
\put ( 37.75,122.5) {82(123)}
\put ( 47.75,115) {\vector(-1,-4){10.25}}

\put ( 41.5 ,190) {\line(-1,-4){6}}
\put ( 20.5 ,153.5) {16(225)}
\put ( 30.5 ,146) {\vector(-1,-4){18}}

\put (250,  0) {$N(939)P_{11}$}
\put (250, 73) {$\Delta(1232)P_{33}$}
\put (250,135) {$N(1520)D_{13}$}
\put (250,150) {$N(1535)S_{11}$}
\put (250,165) {$N(1650)S_{11}$}
\put (250,180) {$N(1675)D_{15}$}
\put (250,195) {$N(1700)D_{13}$}

\end{picture}
\vspace{15pt}
\caption[]{\small 
$N \pi$ and $\Delta \pi$ decay widths of negative parity nucleon
resonances with $(v_1,v_2),L^P=(0,0),1^-$.  
The theoretical values are in parenthesis. All values in MeV.
\normalsize} 
\label{nwidths}
\end{figure}
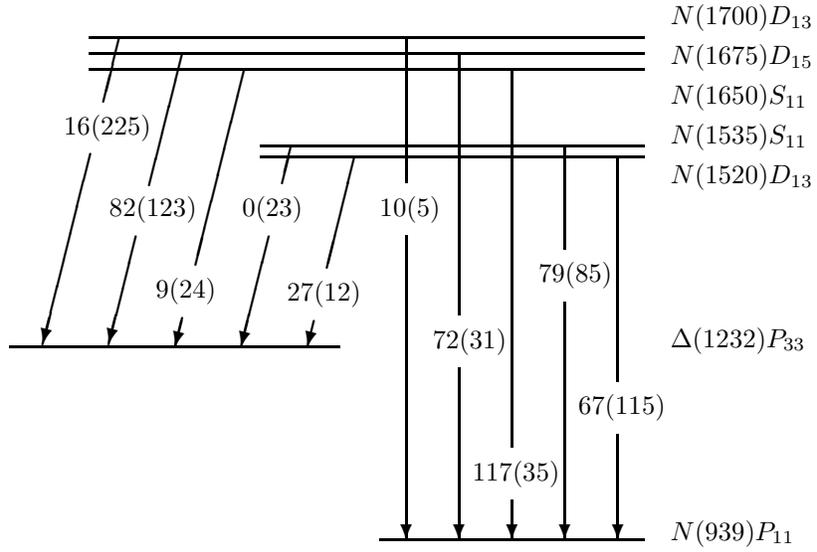

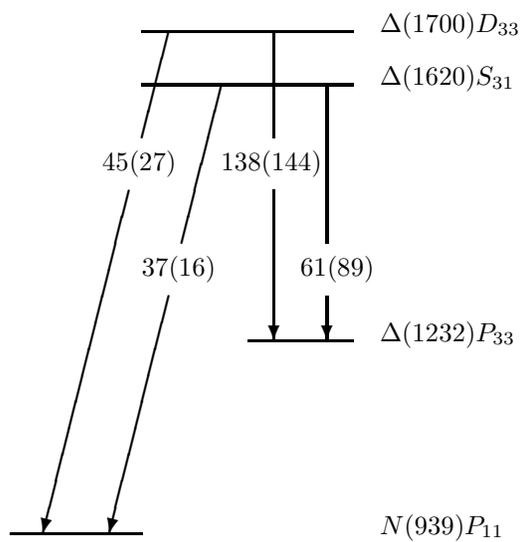
\begin{figure}
\centering
\setlength{\unitlength}{1pt}
\begin{picture}(250,250)(0,0)
\thicklines
\put (  0,  0) {\line(1,0){ 50}}
\put ( 90, 73) {\line(1,0){ 40}}
\put ( 50,170) {\line(1,0){ 80}}
\put ( 50,190) {\line(1,0){ 80}}

\put (120,170) {\line( 0,-1){60}}
\put (110, 97.5) {61(89)}
\put (120, 90) {\vector( 0,-1){17}}

\put (100,190) {\line( 0,-1){40}}
\put ( 80,137.5) {138(144)}
\put (100,130) {\vector( 0,-1){57}}

\put (80,170) {\line(-1,-4){15}}
\put (50, 97.5) {37(16)}
\put (60, 90) {\vector(-1,-4){22.5}}

\put (60,190) {\line(-1,-4){10}}
\put (35,137.5) {45(27)}
\put (45,130) {\vector(-1,-4){32.5}}

\put (140,  0) {$N(939)P_{11}$}
\put (140, 73) {$\Delta(1232)P_{33}$}
\put (140,170) {$\Delta(1620)S_{31}$}
\put (140,190) {$\Delta(1700)D_{33}$}

\end{picture}
\vspace{15pt}
\caption[]{\small 
$N \pi$ and $\Delta \pi$ decay widths of negative parity delta 
resonances with $(v_1,v_2),L^P=(0,0),1^-$.  
The theoretical values are in parenthesis. All values in MeV.
\normalsize} 
\label{dwidths}
\end{figure}

The calculation of decay widths into the $N \pi$ channel, 
as shown for the 3 and 4 star resonances in Table~\ref{npi}, is in 
fair agreement with experiment. This is emphasized in 
Figures~\ref{nwidths} and~\ref{dwidths}, where we plot the $N \pi$ 
and $\Delta \pi$ decay widths of the negative parity resonances 
of Table~\ref{negwidth}. The results are to a large 
extent a consequence of spin-flavor symmetry. The use of collective 
form factors improves somewhat the results when compared with 
harmonic oscillator calculations. This is shown in Table~\ref{regge} 
where the decay of a $\Delta$ Regge trajectory into $N \pi$ 
is analyzed and compared with the calculations of \cite{LeYaouanc}, 
which are based on the harmonic oscillator model discussed in 
\cite{Dalitz}. We also include the results of 
more recent calculations in the nonrelativistic 
quark model \cite{KI} and in the relativized quark model \cite{CR}.

Contrary to the decays into $\pi$, the decay widths into $\eta$ have 
some unusual properties. Our calculation gives systematically small 
values for these widths. This is due to a combination of phase space 
factors and the structure of the transition operator. 
The spin-flavor part is approximately the same for $N \pi$ and $N \eta$, 
since $\pi$ and $\eta$ are in the same $SU_f(3)$ multiplet. 
The $\eta$ decays are suppressed relative to the $\pi$ decays 
because of phase space (due to the difference between the $\eta$ and 
$\pi$ masses). 
Moreover, for small values of $ka$ the elementary amplitudes of 
Eqs.~(\ref{pw02}) are dominated by the $S$ wave amplitude, 
$W_0(k) \approx 3h m_3 k_0 a$, whose contribution to the $\eta$ decay
width is suppressed by the small value of $h=-0.094$ GeV$^{-1}$.
We emphasize here, that 
the structure of the transition operator was determined 
by fitting the coefficients $g$ and $h$ to the $N \pi$ decays 
of the 3 and 4 star resonances. Hence the $\eta$ decays are 
calculated without introducing any further parameters. 

The experimental situation is unclear. The previous 
PDG compilation \cite{PDG92} gave systematically small widths 
($\sim 1$ MeV) for all resonances except $N(1535)S_{11}$. 
The latest compilation deletes all $\eta$ widths with the 
exception of $N(1535)S_{11}$. The results of our analysis 
suggest that the large $\eta$ width for $N(1535)S_{11}$ is not due 
to a conventional $q^3$ state. One possible explanation is 
the presence of another state in the same mass region, {\it e.g.} 
a quasi-bound meson-baryon $S$ wave resonance, just below or around 
threshold. Recently it was suggested \cite{Kaiser}
that a quasi-bound $K \Sigma$-$K \Lambda$ state with properties 
remarkably similar to the $N(1535)S_{11}$ resonance could be 
responsable for the large $N\eta$ branch of the $N(1535)S_{11}$ 
resonance.

\section{Conclusions}

We have presented a calculation of the strong decay widths of 
nucleon and delta resonances into the $\pi$ and $\eta$ channel.
By exploiting the symmetry of the problem, both in its spin-flavor-color 
part, $SU_{sf}(6) \otimes SU_c(3)$, and in its space part, $U(7)$, we 
have been able to express the results in a transparent analytic way.
The analysis of experimental data shows that, while the decays into 
$\pi$ follow the expected pattern, the decays into $\eta$ have some 
unusual features. Our calculations do not show any indication for a 
large $\eta$ width, as is observed for the $N(1535)S_{11}$ resonance. 
The observed large $\eta$ width indicates the presence of another 
configuration, which is outside the present model space. 
A possible candidate for such a configuration is a quasi-bound 
meson-baryon $S$ wave resonance ($K \Sigma$-$K \Lambda$ ) \cite{Kaiser}. 
Experiments at the new electron facilities (CEBAF, MAMI, ELSA) 
could help to elucidate this point.

Other decay channels, such as $\Lambda K$ and 
$\Sigma K$, are presently being included in our calculations.  
This work forms part of a study to incorporate strange resonances 
as well in our model.

\section*{Acknowledgements}

This work is supported in part
by CONACyT, M\'exico under project 400340-5-3401E, DGAPA-UNAM under
project IN105194 (R.B.), and 
by grant No. 94-00059 from the United States-Israel Binational Science
Foundation (BSF), Jerusalem, Israel (A.L.).

\clearpage
\begin{table}
\centering
\caption[]{\small
Matrix elements ${\cal F}(k)$ in the collective 
model for $N \rightarrow \infty$ (large model space). 
The final state is $[56,0^+]_{(0,0);0}$. 
\normalsize}
\label{collff} \vspace{15pt}
\begin{tabular}{cc}
\hline
& \\
Initial state & ${\cal F}(k)$ \\
& \\
\hline
& \\
$[56,0^+]_{(0,0);0}$ 
& $\frac{1}{(1+k^2a^2)^2}$ \\
& \\
$[20,1^+]_{(0,0);0}$ & 0 \\
& \\
$[70,1^-]_{(0,0);1}$ & $i \, \sqrt{3} \, \frac{ka}{(1+k^2a^2)^2}$ \\
& \\
$[56,2^+]_{(0,0);0}$ 
& $\frac{1}{2}\sqrt{5}\left[ \frac{-1}{(1+k^2a^2)^2} + \frac{3}{2k^3a^3} 
\left( \arctan ka - \frac{ka}{1+k^2a^2} \right) \right]$ \\
& \\
$[70,2^-]_{(0,0);1}$ & 0 \\
& \\
$[70,2^+]_{(0,0);2}$ 
& $-\frac{1}{2}\sqrt{15}\left[ \frac{-1}{(1+k^2a^2)^2} + \frac{3}{2k^3a^3} 
\left( \arctan ka - \frac{ka}{1+k^2a^2} \right) \right]$ \\
& \\
\hline
\end{tabular}
\end{table}

\clearpage
\begin{table}
\centering
\caption[]{\small 
Coefficients $c_l$ of Eq.~(\ref{width}) for the strong decay widths 
$\Gamma(B \rightarrow B^{\prime}+M)$ of the negative parity resonances 
with $(v_1,v_2),L^P=(0,0),1^-$. The final state is 
$^{2}8_{1/2}[56,0^+]_{(0,0);0}$ for the nucleon ($B^{\prime}=N$) and 
$^{4}10_{3/2}[56,0^+]_{(0,0);0}$ for the delta ($B^{\prime}=\Delta$). 
The coefficient 
$\xi=(\cos \theta_P - \sqrt{2} \sin \theta_P)/\sqrt{3}$ for $M=\eta$ and
$\xi=(\sin \theta_P + \sqrt{2} \cos \theta_P)/\sqrt{3}$
for $M=\eta^{\prime}$. 
\normalsize}
\label{negwidth}
\vspace{15pt}
\begin{tabular}{cc|cc|cc|cc|cc}
\hline
& & & & & & & & & \\
\multicolumn{2}{c|} {State} & \multicolumn{2}{c|} {$N \pi$} 
& \multicolumn{2}{c|} {$N \eta$}
& \multicolumn{2}{c|} {$\Delta \pi$}  
& \multicolumn{2}{c} {$\Delta \eta$} \\
& & $c_0$ & $c_2$ & $c_0$ & $c_2$ & $c_0$ & $c_2$ & $c_0$ & $c_2$ \\
& & & & & & & & & \\
\hline
& & & & & & & & & \\
$S_{11}$ & $^{2}8_{1/2}[70,1^-]_{(0,0);1}$ 
& $\frac{8}{3}$ & & $2 \xi^2$ & & & $\frac{16}{3}$ & & \\
& & & & & & & & & \\
$D_{13}$ & $^{2}8_{3/2}[70,1^-]_{(0,0);1}$ 
& & $\frac{8}{3}$ & & $2 \xi^2$ & $\frac{8}{3}$ & $\frac{8}{3}$ & & \\
& & & & & & & & & \\
$S_{11}$ & $^{4}8_{1/2}[70,1^-]_{(0,0);1}$ 
& $\frac{2}{3}$ & & $2 \xi^2$ & & & $\frac{4}{3}$ & & \\
& & & & & & & & & \\
$D_{13}$ & $^{4}8_{3/2}[70,1^-]_{(0,0);1}$ 
& & $\frac{1}{15}$ & & $\frac{1}{5} \xi^2$ & $\frac{20}{3}$ 
& $\frac{64}{15}$ & & \\
& & & & & & & & & \\
$D_{15}$ & $^{4}8_{5/2}[70,1^-]_{(0,0);1}$ 
& & $\frac{2}{5}$ & & $\frac{6}{5} \xi^2$ & & $\frac{28}{5}$ & & \\
& & & & & & & & & \\
\hline
& & & & & & & & & \\
$S_{31}$ & $^{2}10_{1/2}[70,1^-]_{(0,0);1}$ 
& $\frac{1}{3}$ & & & & & $\frac{20}{3}$ & & $4 \xi^2$ \\
& & & & & & & & & \\
$D_{33}$ & $^{2}10_{3/2}[70,1^-]_{(0,0);1}$ 
& & $\frac{1}{3}$ & & & $\frac{10}{3}$ & $\frac{10}{3}$ 
& $2 \xi^2$ & $2 \xi^2$ \\
& & & & & & & & & \\
\hline
\end{tabular}
\end{table}

\clearpage
\begin{table}
\centering
\caption[]{\small 
$N \pi$, $N \eta$ and $\Delta \eta$ decay widths of (3 and 4 star) 
nucleon and delta 
resonances in MeV. The experimental values are taken from \cite{PDG}.
The mixing angle for the $\eta$ mesons is $\theta_P=-23^{\circ}$ 
\cite{GIK}.
\normalsize}
\label{npi} 
\vspace{15pt} 
\begin{tabular}{lcccccccc}
\hline
& & & & & & & & \\
State & Mass & Resonance & \multicolumn{2}{c}{$\Gamma(N \pi)$} 
& \multicolumn{2}{c}{$\Gamma(N \eta)$} 
& \multicolumn{2}{c}{$\Gamma(\Delta \eta)$} \\
& & & th & exp & th & exp & th & exp \\
& & & & & & & & \\
\hline
& & & & & & & & \\
$S_{11}$ & $N(1535)$ & $^{2}8_{1/2}[70,1^-]_{(0,0);1}$ 
&  $85$ & $79 \pm 38$ &  $0.1$ & $86 \pm 85$ & & \\
$S_{11}$ & $N(1650)$ & $^{4}8_{1/2}[70,1^-]_{(0,0);1}$ 
&  $35$ & $117 \pm 23$ &  $8$   & & & \\
$P_{13}$ & $N(1720)$ & $^{2}8_{3/2}[56,2^+]_{(0,0);0}$ 
&  $31$ & $22 \pm 11$ &  $0.2$ & & & \\
$D_{13}$ & $N(1520)$ & $^{2}8_{3/2}[70,1^-]_{(0,0);1}$ 
& $115$ & $67 \pm 9$ &  $0.6$ & & & \\
$D_{13}$ & $N(1700)$ & $^{4}8_{3/2}[70,1^-]_{(0,0);1}$ 
&   $5$ & $10 \pm 7$ &  $4$   & & & \\
$D_{15}$ & $N(1675)$ & $^{4}8_{5/2}[70,1^-]_{(0,0);1}$ 
&  $31$ & $72 \pm 12$ & $17$   & & & \\
$F_{15}$ & $N(1680)$ & $^{2}8_{5/2}[56,2^+]_{(0,0);0}$ 
&  $41$ & $84 \pm 9$ &  $0.5$ & & & \\
$G_{17}$ & $N(2190)$ & $^{2}8_{7/2}[70,3^-]_{(0,0);1}$ 
&  $34$ & $67 \pm 27$ & $11$   & & & \\
$G_{19}$ & $N(2250)$ & $^{4}8_{9/2}[70,3^-]_{(0,0);1}$ 
&  $7$ & $38 \pm 21$ &  $9$   & & & \\
$H_{19}$ & $N(2220)$ & $^{2}8_{9/2}[56,4^+]_{(0,0);0}$ 
&  $15$ & $65 \pm 28$ &  $0.7$ & & & \\
$I_{1,11}$ & $N(2600)$ & $^{2}8_{11/2}[70,5^-]_{(0,0);1}$ 
&  $9$ & $49 \pm 20$ & $3$ & & & \\
& & & & & & & & \\
\hline
& & & & & & & & \\
$S_{31}$ & $\Delta(1620)$ & $^{2}10_{1/2}[70,1^-]_{(0,0);1}$ 
&  $16$ & $37 \pm 11$ & & & $-$ & $-$ \\
$P_{31}$ & $\Delta(1910)$ & $^{4}10_{1/2}[56,2^+]_{(0,0);0}$ 
&  $42$ & $52 \pm 19$ & & & $0.0$ & \\
$P_{33}$ & $\Delta(1232)$ & $^{4}10_{3/2}[56,0^+]_{(0,0);0}$ 
& $116$ & $119 \pm 5$ & & & $-$ & $-$ \\
$P_{33}$ & $\Delta(1920)$ & $^{4}10_{3/2}[56,2^+]_{(0,0);0}$ 
&  $22$ & $28 \pm 19$ & & & $0.5$ & \\
$D_{33}$ & $\Delta(1700)$ & $^{2}10_{3/2}[70,1^-]_{(0,0);1}$ 
&  $27$ & $45 \pm 21$ & & & $-$ & $-$ \\
$D_{35}$ & $\Delta(1930)$ & $^{2}10_{5/2}[70,2^-]_{(0,0);1}$ 
&   $0$ & $52 \pm 23$ & & & $0$   & \\
$F_{35}$ & $\Delta(1905)$ & $^{4}10_{5/2}[56,2^+]_{(0,0);0}$ 
&   $9$ & $36 \pm 20$ & & & $1$   & \\
$F_{37}$ & $\Delta(1950)$ & $^{4}10_{7/2}[56,2^+]_{(0,0);2}$ 
&  $45$ & $120 \pm 14$ & & & $2$   & \\
$H_{3,11}$ & $\Delta(2420)$ & $^{4}10_{11/2}[56,4^+]_{(0,0);0}$ 
&  $12$ & $40 \pm 22$ & & & $2$ & \\
& & & & & & & & \\
\hline
\end{tabular}
\end{table}

\clearpage
\begin{table}
\centering
\caption[]{\small 
Strong decay widths for $\Delta^{\ast} \rightarrow N + \pi$ 
and $N^{\ast} \rightarrow N + \pi$ in MeV.
Experimental values are from \cite{PDG}.
\normalsize}
\label{regge} 
\vspace{15pt} 
\begin{tabular}{lcccccc}
\hline
& & & & & & \\
Resonance & $L$ & \multicolumn{4}{c} {$\Gamma$(th)} & $\Gamma$(exp) \\
& & Ref.~\cite{LeYaouanc} & Ref.~\cite{KI} & Ref.~\cite{CR} & Present & \\
& & & & & & \\
\hline
& & & & & & \\
$\Delta(1232)P_{33}$   & 0 & 70 & 121 & 108 & 116 & $119 \pm 5 $ \\
$\Delta(1950)F_{37}$   & 2 & 27 &  56 &  50 &  45 & $120 \pm 14$ \\
$\Delta(2420)H_{3,11}$ & 4 &  4 &     &   8 &  12 & $ 40 \pm 22$ \\
$\Delta(2950)K_{3,15}$ & 6 &  1 &     &   3 &   5 & $ 13 \pm 8 $ \\
& & & & & & \\
\hline
& & & & & & \\
$N(1520)D_{13}$   & 1 & & 85 & 74 & 115 & $67 \pm 9 $ \\
$N(2190)G_{17}$   & 3 & &    & 48 &  34 & $67 \pm 27$ \\
$N(2600)I_{1,11}$ & 5 & &    & 11 &   9 & $49 \pm 20$ \\
& & & & & & \\
\hline
\end{tabular}
\end{table}

\clearpage

\clearpage

\clearpage

\clearpage

\end{document}